\RequirePackage{fix-cm}

\documentclass[twocolumn]{svjour3}

\usepackage{graphicx}
\usepackage{mathptmx}     
\usepackage{latexsym}
\usepackage{doi}
\usepackage[normalem]{ulem}

\usepackage{amsmath,mathrsfs,amssymb}
\usepackage{physics}
\usepackage{tensor}
\usepackage{geometry}
\usepackage{xcolor}
\usepackage{enumitem}
\usepackage{tabularx}

\begin{document}

\title{Determination of the metric of a Schwarzschild black hole surrounded by a halo of dark matter}
%\subtitle{}

\author{M. Castillo Alarcón  \and
       Leonel Bixano \and  I. A. Sarmiento-Alvarado \and Claudio Salas-P\'erez \and Tonatiuh Matos
}
%\affiliation{Departamento de F\'{\i}sica, Centro de Investigaci\'on y de Estudios Avanzados del Instituto Politécnico Nacional, Av. Instituto Politécnico Nacional 2508, San Pedro Zacatenco, M\'exico 07360, CDMX. }%

\institute{%F. Author \at
              Departamento de F\'{\i}sica, Centro de Investigaci\'on y de Estudios Avanzados del Instituto Politécnico Nacional, Av. Instituto Politécnico Nacional 2508, San Pedro Zacatenco, M\'exico 07360, CDMX. \\
              \email{mayte.castillo@cinvestav.mx, 
              leonel.delacruz@cinvestav.mx,
              ignacio.sarmiento@cinvestav.mx,
              claudio.salas@cinvestav.mx,
              tonatiuh.matos@cinvestav.mx}      
}

\date{}

\maketitle

\begin{abstract}
Following the landmark acquisition of the first image of a black hole, systems comprising a black hole enveloped by a dark matter halo have attracted considerable attention. This work implements a novel method to analyze the structure of a system composed of a Schwarzschild black hole and a dark matter halo modeled as an anisotropic fluid, while ensuring that the proposed metric constitutes an exact solution of the Einstein field equations. Numerical results show that the connection between the temporal and radial metric functions preserves an inverse relationship, namely $A(r) \approx 1/B(r)$. Furthermore, we analytically demonstrate that the resulting functions satisfy $A(r)B(r) = 1 + \mathcal{O}(10^{-6})$, in excellent agreement with the numerical findings. In addition, the Einstein equations imply that the variation of the product \(A(r)B(r)\) is governed by the factor \(\varepsilon_h=\frac{8\pi G\rho_0r_0^2}{c^2}\). Finally, we show that the interior Schwarzschild region and the exterior dark matter halo can be joined smoothly via a regular Darmois–Israel junction, without generating any thin-shell curvature steps. To assess the observational implications of the model, we compare its theoretical predictions for the pericenter precession of the S2 star and the angular size of the shadow of Sgr A*. In both cases, the predicted deviations remain within the observational uncertainties reported by the GRAVITY Collaboration and the Event Horizon Telescope (EHT), respectively. These results demonstrate that the proposed black hole–dark matter halo configuration is not only mathematically consistent, but also compatible with current observational constraints.

\end{abstract}

\section{Introduction}
\label{intro}

Dark matter is a central open problem in modern astrophysics and cosmology. Its microscopic nature is unknown, but its gravitational effects are inferred from galactic rotation curves, gravitational lensing, galaxy-cluster dynamics, and large-scale structure. At galactic scales, these observations and simulations have motivated several phenomenological density profiles, notably the cuspy Navarro-Frenk-White profile \cite{Navarro:1996gj}, the cored Burkert profile \cite{Burkert:1995}, and the Einasto profile, whose continuously varying logarithmic slope fits simulated halos well \cite{Graham:2005xx}. Alternative microscopic models, such as scalar-field dark matter \cite{Matos:1998vk}, also yield self-gravitating halos, including multistate configurations compatible with observed galactic rotation curves \cite{Matos:1999et,Urena-Lopez:2010zva}, VPOs \cite{Bernal:2024} and magnetic fields \cite{Hernandez-Marquez:2026lkw}.

Most of these profiles were originally formulated in a Newtonian or weak-field regime and without an explicit central compact object. Yet supermassive black holes are expected at the centers of massive galaxies. Horizon-scale observations of M87 and Sagittarius A* by the Event Horizon Telescope \cite{EventHorizonTelescope:2019dse,EventHorizonTelescope:2022wkp}, together with precise measurements of S-star orbits at the Galactic center \cite{GRAVITY:2020gka,GRAVITY:2024tth}, motivate relativistic models that self-consistently describe a central black hole and its surrounding matter.

%%   \textcolor{blue}{\sout{The dark-matter configuration near a black hole need not follow the galaxy’s large-scale profile. Slow, adiabatic black-hole growth can generate a steep density spike \cite{Gondolo:1999ef}, and relativistic studies show that strong gravity and spin modify the innermost distribution \cite{Sadeghian:2013laa,Ferrer:2017xwm}. Thus, extending a galactic profile down to the event horizon is best viewed as an effective, stationary modeling choice, not the result of a fully dynamical formation scenario.}}

%

Several methods have been proposed to model black holes in dark-matter halos. In Ref \cite{Xu:2018wow}, the halo metric was inferred from a Newtonian rotation curve and then supplemented with a central black-hole term. Although such metrics are widely used for shadows, lensing, quasinormal modes, and particle motion, a Newtonian circular-velocity profile does not uniquely determine a relativistic spacetime, so the resulting geometry must be independently checked against the full Einstein equations. Fully relativistic black-hole halo configurations with anisotropic matter have been constructed: Cardoso et al \cite{Cardoso:2021wlq} obtained an asymptotically flat exact solution using a Hernquist-inspired Einstein cluster, and subsequent work developed more general numerical and analytical models for Hernquist, Navarro Frenk White, Einasto, core, and cusp profiles \cite{Konoplya:2022hbl,Figueiredo:2023gas,Shen:2023erj}. These studies also showed that the density behavior near the horizon and the presence of an inner halo radius are crucial for satisfying the energy conditions \cite{Shen:2024qbb}.

A density profile alone does not fully specify a relativistic source: one must also analyze the radial and tangential pressures, conservation equations, energy conditions, and effective propagation speeds. Some common halo metrics violate the dominant energy condition or yield pathological effective sound speeds \cite{Datta:2023zmd}. It has furthermore been shown that simultaneously imposing a Newtonian rotation-curve prescription and \(g_{tt}=1/g_{rr}\) generically leads to an anisotropic source with \(P_r=-c^2\rho\), whose density can differ from the assumed halo profile \cite{Bolokhov:2025zva}. Thus, the relation between the metric functions and the effective equation of state must be derived from the Einstein equations rather than imposed independently.

%% \textcolor{blue}{\sout{We consider a static, spherically symmetric line element \(ds^2=-A(r)c^2dt^{2}+B(r)dr^{2}+r^{2}d\Omega^{2}\), with \(A(r)\) and \(B(r)\) initially independent. Instead of using rotation curves, we insert the density directly into the Einstein equations, obtaining the metric functions and anisotropic pressures from the full relativistic system. This yields an effective stationary description where the halo profile is prescribed rather than derived from black-hole formation, accretion, capture, or relaxation, while still including its gravitational deformation through the Einstein equations. We apply this method to Burkert, Einasto, and multistate scalar field dark matter profiles, thus comparing phenomenological cored distributions, simulation-motivated profiles, and configurations from a microscopic bosonic model.}}

In this article, we present a model of a Schwarzschild black hole in a dark matter halo that is derived directly from Einstein equations. We consider a static, spherically symmetric line element \(ds^2=-A(r)c^2dt^{2}+B(r)dr^{2}+r^{2}d\Omega^{2}\); unlike other models, we assume that \(A(r)\) and \(B(r)\) are initially independent. In doing so, we respect the symmetries imposed by the dark matter halo and Schwarzschild black hole models without restricting the metric functions to the case where the halo is not present $A(r)B(r)=1$.

Instead of using rotation curves, we introduce the density profile of dark matter directly into Einstein equations, modeling the energy-momentum tensor of dark matter as an anisotropic fluid, unlike previous work in which $P_r=P_T$ or $P_r=0$ is imposed. This leads to an effective stationary description in which the halo profile is set a priori, rather than being derived from the formation, accretion, capture, or relaxation of the black hole, while its gravitational deformation is accounted for via Einstein equations. We apply this framework to the Burkert, Einasto, and multistate scalar field dark matter profiles, demonstrating that the formalism is applicable to phenomenological, simulation-motivated, and microscopic dark matter models.

In this way, both the metric functions and the anisotropic pressures are obtained from the complete relativistic system by analytically solving Einstein equations, without any of them being imposed independently. We also present a numerical analysis that supports the analytical results, which together fully justify the assumptions regarding the metric functions and the equations of state for dark matter that are imposed in other works as a convenient ansatz.

The structure of the article is as follows. In section \ref{sec:2}, we construct the metric on which our method is based and solve Einstein’s equations exactly. In section \ref{sec:3}, we derive the radial and tangential pressures of dark matter and verify the strong energy condition; furthermore, we analytically demonstrate the relationship between the metric functions obtained in section 2 and show that the Darmois-Israel junction conditions are satisfied. Finally, in Section \ref{sec:4}, we present the results obtained for the three density profiles and we present the observational constraints.

%..........................................
\section{The geometry of the model}
\label{sec:2}
We assume spherical symmetry for dark matter halo and Schwarzschild black hole separately, then we adopt a metric of the form
{\footnotesize
\begin{equation}\label{eq:metric_AB}
ds^2 = -A(r)\,c^2dt^2 + B(r)\,dr^2 + r^2\big(d\theta^2+\sin^2\theta\,d\phi^2\big).
\end{equation}
}
The main difference from the most commonly used hypothesis in the literature \cite{xu2018}, in which $A(r)=B(r)^{-1}$, is that we do not assume any relationship between the metric functions.\\ 
On the other hand, with the intention of removing restrictions on the dark matter model we will choose the energy-momentum tensor corresponding to an anisotropic fluid defined by
\begin{equation}\label{eq:T_Anisotropico}
T^\mu {}_\nu = \mathrm{diag}\big(-c^2\varrho(r),\;P_r(r),\;P_T(r),\;P_T(r)\big),
\end{equation}
where $P_r(r)\neq P_T(r)$, in contrast with the model in \cite{Cardoso2022}. This configuration must satisfy the Einstein equations $G^\mu{}_\nu = \frac{8\pi G}{c^4}\,T^\mu{}_\nu$, explicitly, we have the system of equations. 
{\footnotesize
\begin{subequations} 
\begin{align}
\frac{8\pi G}{c^2} \varrho(r) &=\frac{B'}{B^2\,r}+\frac{1-\frac1B}{r^2},\label{eq:rho_AB} \\
\frac{8\pi G}{c^4} P_r(r) &=\frac{A'}{A\,B\,r}+\frac{\frac1B-1}{r^2},\label{eq:Pr_AB} \\
\frac{8\pi G}{c^4} P_T(r) &=
\frac{A''}{2AB}
-\frac{A'B'}{4AB^2}
-\frac{(A')^2}{4A^2B}
-\frac{B'}{2rB^2}
+\frac{A'}{2rAB}.\label{eq:Pt_AB}
\end{align}
\end{subequations}
}
where primes denote $\dv{}{r}$.

\subsection{Determination of $B(r)$}
\label{sec:2.1}

By analogy with the solution to Einstein equations in a vacuum in a spherically symmetric configuration, in this paper we will assume that
\begin{equation}\label{eq:B-ansatz}
    B(r)=\left(1-\frac{2Gm(r)}{c^2r}\right)^{-1},
\end{equation}
employing this ansatz in equation \eqref{eq:rho_AB}, we find
\begin{equation}\label{eq:mprima}
    m'(r)=4\pi r^2 \varrho(r),\quad r\geq r_{\mathrm{in}}
\end{equation}
where \(\rho(r)=\Theta(r-r_{in})\,\rho_h(r)\), with \( \Theta \) denoting the Heaviside step function, and 
\begin{equation}\label{eq:Def_rin}
    r_{in}:=\frac{6GM_\bullet}{c^2}.
\end{equation}
We will assume that the density of dark matter is nonzero only if $r>r_{in}$, which is the boundary beyond which a geodesic of a massive particle is stable in the Schwarzschild black hole. This means that the contribution of dark matter to the system for $r<r_{in}$ is zero, since all dark matter located within this radial threshold is captured by the black hole. This assumption is reflected in the definition of the dark matter density using the Heaviside step function. A more realistic description of the system must include a dynamic model of dark matter for $r<r_{in}$.

Integrating we have
\begin{equation}\label{eq:m_quadrature}
m(r)=M_\bullet + m_h(r)=M_\bullet + 4\pi\int_{r_{in}}^{r}\tilde r^2\,\varrho(\tilde r)\,\dd \tilde r,
\end{equation}
where 
\begin{equation}\label{eq:mh_definition}
    m_h(r)=
    \begin{cases}
        0,
        &
        r\leq r_{\mathrm{in}},\\
        4\pi
        \int_{r_{\mathrm{in}}}^{r}
        \tilde r^{\; 2}
        \rho_h(\tilde r),
        \mathrm{d}\tilde r,
        &
        r>r_{\mathrm{in}}.
    \end{cases}
\end{equation}

The integration constant \(M_\bullet\) is identified with the mass of the black hole, and \(m_h(r)\) with the halo mass. Thus, the metric function $B(r)$ takes the form
\begin{equation}\label{eq:B-with-halo}
    B(r)=\left(1-\frac{2GM_\bullet}{c^2r}-\frac{2G m_h(r)}{c^2r}\right)^{-1}.
\end{equation}
It is easy to see that if we consider the case of black hole only ($m_h(r)=0, \quad r<r_{in}$) we recover the metric function corresponding to Schwarzschild solution.

\subsection{Determination of $A(r)$}
\label{sec:2.2}

For the other metric function $A(r)$, taking into account the approximation in the weak-field metric, we will decompose the gravitational potential into a contribution from the black hole alone and a deformation generated by the halo, using the ansatz
\begin{equation}\label{eq:A-ansatz}
    A(r)=\left(1-\frac{2GM_\bullet}{c^2r}\right)e^{\frac{2\psi(r)}{c^2}}.
\end{equation}
The geometric part of the model is thus completely determined by the two assumptions in the metric functions \eqref{eq:B-ansatz} and \eqref{eq:A-ansatz}; note that:
\begin{enumerate}[label=(\alph*)]
    \item We recover Schwarzschild solution exactly if the halo is absent, i.e. $m_h=0$ then the exponential function $\psi=0$.
    \item All halo information entering the radial sector is already contained in $m_h(r)$.
    \item The function $\psi(r)$ cleanly measures the halo induced deformation.
\end{enumerate}

From the proposed function \eqref{eq:A-ansatz}, we can derive the relation  
\begin{equation}\label{eq:Aprime-over-A-final}
    \frac{A'}{A}=\frac{2GM_\bullet}{r(c^2r-2GM_\bullet)}+\frac{2\psi'(r)}{c^2}.
\end{equation}

By substituting \eqref{eq:B-with-halo} and \eqref{eq:Aprime-over-A-final} into the Einstein equation \eqref{eq:Pr_AB} we derive the differential equation for the deformation function $\psi(r)$
\begin{equation}\label{eq:psi-pr-general}
    \psi'(r)=\frac{4\pi G r^2P_r(r)+\dfrac{Gc^4m_h(r)}{c^2r-2GM_\bullet}}{c^2r-2GM_\bullet-2Gm_h(r)}.
\end{equation}
In other words, when we consider the vacuum case with \(m_h=0=P_r\), we can assume that the deformation function becomes \(\psi=0\), and the solution reduces exactly to the Schwarzschild metric, just as we expected. In the other case, we need to calculate the deformation induced by the halo using \eqref{eq:psi-pr-general}, which means we need the exact form of the pressure $P_r(r)$, we will determine this in the next section.

\section{Dark matter characterization}
\label{sec:3}
\subsection{Determination of $P_r$}
\label{sec:3.1}

It is well established that the Schwarzschild solution represents a vacuum solution of Einstein field equations, meaning that both pressure and density vanish. Consequently, when this fact is taken into consideration, the pressure in the composite configuration we are examining, consisting of dark matter and a central a Schwarzschild black hole, will be predominantly determined by the dark matter halo. In this way, we can obtain the expression for the radial pressure \(P_r\) from the tangential velocities of the stars in the galaxy, which are modeled without considering the black hole at its center.\\
Then we consider the halo on its own, for this case, we introduce a halo mass function \(m_h(r)\) determined by \eqref{eq:m_quadrature} with \(M_\bullet = 0\), and we define the radial metric component as
\begin{equation*}
    B_h(r) = \left(1 - \frac{2Gm_h(r)}{c^2r}\right)^{-1}.
\end{equation*}
And the temporal metric function takes the form
 \begin{equation*}
      A_h(r) = e^{\frac{2\psi_h(r)}{c^2}},
 \end{equation*}
 where $\psi_h(r)$ is the deformation function, which does not take into account the black hole, which can be identified as the gravitational potential. If circular geodesics in the halo alone exhibit an observed tangential velocity
 \begin{equation}\label{eq:velocitytg}
     v_c^2(r)=\frac{Gm_h(r)}{r},
 \end{equation}
 the usual exact relation between circular orbits and the gravitational potential implies
\begin{equation}\label{eq:Velocidad Circular}
    q(r):=\frac{v_c^2(r)}{c^2}=\frac{r\psi_h'(r)}{c^2}=\frac{r}{2}\frac{A_h'(r)}{A_h(r)}.
\end{equation}
Thus, by applying the identities \eqref{eq:Aprime-over-A-final} and \eqref{eq:psi-pr-general} given that $M_\bullet=0$, we find
{\small
\begin{equation}\label{eq:Pr-hat-definition}
4\pi G r^2 \widehat P_r(r)=q(r)c^4\left(1-\frac{2Gm_h(r)}{c^2r}\right)-\frac{Gc^2m_h(r)}{r}.
\end{equation}
}

Here, $\widehat{P}_r(r)$ represents the radial pressure in the case of a halo-only configuration. The model approximation introduced in this study is given by
\begin{equation}\label{eq:single-approximation}
    P_r(r)=\widehat P_r(r).
\end{equation}
In other words, the exact radial pressure of the full system, comprising both the black hole and the dark matter halo, is set equal to the radial pressure that is reconstructed using only the halo component. This approach is less restrictive than the one used in \cite{Cardoso2022} and \cite{Shen:2023}, where it is assumed that $P_r(r)=0$ from the outset. \\
Consequently, the deformation function can be expressed as:
{\small
\begin{equation}\label{eq:psi-prime-master}
    \psi'(r)=\frac{q(r)c^2\left[c^2r-2Gm_h(r)\right]+Gc^2m_h(r)\frac{2GM_\bullet}{c^2r-2GM_\bullet}}{r\left[c^2r-2GM_\bullet-2Gm_h(r)\right]}.
\end{equation}
}

\subsection{Determination of $P_T$}
\label{sec:3.2}

The Bianchi identities imply that $\nabla_\mu G^\mu{}_\nu = 0$, which, through the Einstein field equations, yields
\(
    \nabla_\mu T^\mu{}_\nu = 0
\).
For an anisotropic stress-energy tensor, the only nonzero and nontrivial component arises for $\nu = r$, giving rise to the anisotropic Tolman-Oppenheimer-Volkoff relation
\begin{equation}\label{eq:TOV_aniso}
P_r'(r)=-\frac12\frac{A'}{A}\, \left(c^2\rho+P_r\right) + \frac{2}{r}\big(P_T-P_r\big).
\end{equation}
This can be solved algebraically for $P_T$:
\begin{equation}\label{eq:Pt_from_TOV}
    \begin{split}
        P_T(r) &=\; P_r(r) + \frac{r}{2}\bigg[
        P_r'(r)
        + \left(c^2\rho(r) + P_r(r)\right) \\
        &\left(\frac{GM_\bullet}{r(c^2 r - 2GM_\bullet)} + \frac{\psi'(r)}{c^2}\right)
        \bigg],
    \end{split}
\end{equation}

where \(Pr=\hat P_r\). In this way, we have determined the complete system through the density profile $\rho(r)$. Using \eqref{eq:m_quadrature}, we can obtain the radial metric function $B(r)$, and using \eqref{eq:psi-prime-master} and \eqref{eq:velocitytg}, we can calculate the temporal metric function $A(r)$. Furthermore, we can characterize the dark matter model using the pressures determined by \eqref{eq:Pr-hat-definition} and \eqref{eq:Pt_from_TOV}.

Equation \eqref{eq:Pt_from_TOV} specifies the tangential pressure required so that the density and radial pressure together satisfy the anisotropic conservation equation. In this way, \(P_T\) counterbalances the gravitational, radial-pressure, and anisotropic forces in the static configuration. This condition of equilibrium, however, should not be interpreted as implying dynamical stability under radial perturbations. The cracking criterion serves as a useful tool for identifying regions that may become unstable immediately after an anisotropic fluid departs from equilibrium \cite{Herrera:1992,Abreu:2007}, but its local sound-speed formulation presumes constitutive relations of the form \(\delta P_r = v_r^2 \delta\rho\) and \(\delta P_T = v_T^2 \delta\rho\).

In the current model, this condition is not automatically fulfilled because the radial pressure in Eq.~\eqref{eq:Pr-explicito para prueba} depends nonlocally on the density via the enclosed halo mass. In fact, a density perturbation leads to
{\footnotesize
\[
\begin{aligned}
\delta m_h(r)
&=
4\pi\int_{r_{\mathrm{in}}}^{r}
s^2\delta\rho(s)\,\mathrm ds,\
\delta P_r(r)
=
-\frac{Gm_h(r)}{\pi r^4}\,\delta m_h(r).
\end{aligned}
\]
}
Thus, \(\delta P_r(r)\) is influenced by the density perturbation over the entire interior range and, in general, cannot be represented as a local sound speed times \(\delta\rho(r)\). In particular, conservation of the total halo mass only enforces that the integral of \(\delta\rho \) over the whole halo must vanish, requiring \(\delta m_h(r)=0 \) at every radius would instead force \(\delta\rho(r)=0 \) and therefore yield only the trivial perturbation. Furthermore, the perturbation of \(P_T \) also depends on the induced changes in \(P_r' \), \(\psi' \), and the metric functions.

Consequently, the current construction yields a self-consistent static equilibrium configuration, but it neither demonstrates nor rules out dynamical radial stability. A thorough investigation would require specifying a microscopic or constitutive model for the Lagrangian variations of both pressure components, and then solving the linearized Einstein equations together with the conservation laws for the radial normal modes. Carrying out such a dynamical analysis goes beyond the stationary effective framework adopted here.

\subsection{Semi-analytical demonstration of the relationship between the metric functions}\label{SubSec:Analytical proof the the relation between the metric functions}

We now present a semi-analytic demonstration that the metric functions obtained from the Einstein equations satisfy \(A(r)\simeq B(r)^{-1}\), despite this relation not being assumed in the original metric ansatz. The dark matter density is nonvanishing only in the region \(r\geq r_{\mathrm{in}}\), where \(r_{in}\) is given by the definition \eqref{eq:Def_rin}. Consequently, \(m_h(r)=0\) for \(r\leq r_{\mathrm{in}}\), see eq. \eqref{eq:mh_definition}.

Throughout, all equations are assumed to be valid in the domain \(r \geq r_{\mathrm{in}}\). By adding Eqs \eqref{eq:rho_AB} and \eqref{eq:Pr_AB}, and first expressing both left-hand sides with the common factor \(8\pi G/c^4\), we obtain
\[
\frac{8\pi G}{c^4}\left[c^2\varrho(r)+P_r(r)\right]
=
\frac{B'(r)}{B^2(r)r}
+
\frac{A'(r)}{A(r)B(r)r}.
\]
The right-hand side can be rewritten as
\[
\begin{aligned}
    \frac{B'(r)}{B^2(r)r}
    +
    \frac{A'(r)}{A(r)B(r)r}
    &=
    \frac{1}{B(r)r}
    \left[
    \frac{B'(r)}{B(r)}
    +
    \frac{A'(r)}{A(r)}
    \right]
    \\
    & =
    \frac{1}{B(r)r}
    \frac{\mathrm d}{\mathrm dr}
    \ln\!\left[A(r)B(r)\right].
\end{aligned}
\]
Consequently, the Einstein equations imply the exact identity
{\small
\begin{equation}
\frac{\mathrm d}{\mathrm dr}
\ln\!\left[A(r)B(r)\right]
=
\frac{8\pi G}{c^4}B(r)r
\left[c^2\varrho(r)+P_r(r)\right].
\label{eq:AB-indentidad para prueba}
\end{equation}
}

Equation \eqref{eq:AB-indentidad para prueba} demonstrates that \(A(r)B(r)\) would remain strictly constant if the condition \(P_r(r)=-c^2\rho(r)\) were enforced. However, this particular equation of state is not assumed in our model. Consequently, the reciprocal relation between the metric functions is, in general, only approximate. We now proceed to estimate the size of its deviation. For only halo configuration, the tangential velocity is given by \eqref{eq:Velocidad Circular}, then substituting \(q(r)\) onto \eqref{eq:Pr-hat-definition}, we obtain \( 4\pi G r^2P_r(r)=-\frac{2G^2m_h^2(r)}{r^2} \). Then,
\begin{equation}
P_r(r)
=
-\frac{Gm_h^2(r)}{2\pi r^4},
\qquad r\geq r_{\mathrm{in}}.
\label{eq:Pr-explicito para prueba}
\end{equation}
The reconstructed radial pressure takes negative values and depends quadratically on the halo mass. By inserting Eq. \eqref{eq:Pr-explicito para prueba} into Eq. \eqref{eq:AB-indentidad para prueba}, we obtain
{\small
\begin{equation}
\frac{\mathrm d}{\mathrm dr}
\ln\!\left[A(r)B(r)\right]
=
B(r)
\left[
\dfrac{8\pi G}{c^2}r\varrho(r)
-
\frac{4G^2m_h^2(r)}{c^4r^3}
\right].
\label{eq:AB-masas para prueba}
\end{equation}
}
By applying Eq. \eqref{eq:B-with-halo}, this identity can equivalently be written as 
\[
\frac{\mathrm d}{\mathrm dr}
\ln\!\left[A(r)B(r)\right]
=
\frac{
\dfrac{8\pi G}{c^2}r\varrho(r)
-
\dfrac{4G^2m_h^2(r)}{c^4r^3}
}{
1-
\dfrac{2GM_\bullet}{c^2r}
-
\dfrac{2Gm_h(r)}{c^2r}
}.
\]

This equation holds only under the approximation \(P_r=\widehat P_r\). We integrate Eq.~\eqref{eq:AB-masas para prueba} by selecting a reference radius \(r_*>r_{\mathrm{in}}\) and considering \(r\geq r_*\), using \(s\) as the dummy integration variable,
\[
\frac{\mathrm d}{\mathrm ds}
\ln\!\left[A(s)B(s)\right]
=
B(s)
\left[
\frac{2Gm_h'(s)}{c^2s}
-
\frac{4G^2m_h^2(s)}{c^4s^3}
\right].
\]
For the second term, with \(\frac{q^2(s)}{s}=\frac{G^2m_h^2(s)}{c^4s^3}\),  thus,
\[
\frac{\mathrm d}{\mathrm ds}
\ln\!\left[A(s)B(s)\right]
=
\frac{8\pi G}{c^2}B(s)s\rho(s)
-
4B(s)\frac{q^2(s)}{s}.
\]
Integrating from \(r_*\) to \(r\),
\begin{equation}
\begin{aligned}
\ln\!\left[
\frac{A(r)B(r)}{A(r_*)B(r_*)}
\right]
={}&
\frac{8\pi G}{c^2}
\int_{r_*}^{r}B(s)s\rho(s)\,\mathrm ds
\\
&-
4\int_{r_*}^{r}B(s)q^2(s)\frac{\mathrm ds}{s}.
\end{aligned}
\label{eq:AB-integral prueba}
\end{equation}
The first integral is due to the density, and the second to the reconstructed radial pressure.

For \(s>r_{\mathrm{in}}\) we have \(2GM_\bullet/(c^2s)<1/3\). Let \(q_{\max}:=\max_{s\in[r_*,r]}q(s)\), hence \(q_{\max}\ll1\), then
\[
1-\frac{2GM_\bullet}{c^2s}-2q(s)
>
\frac{2}{3}-2q_{\max}
>
0.
\]
Consequently,
\[
1<B(s)\leq B_{\max},
\qquad
B_{\max}:=
\left(\frac{2}{3}-2q_{\max}\right)^{-1}.
\]
Consequently, \(B(s)\) stays finite.

To compute the density integral, we will use \(x := r/r_0\) and \(x_* := r_*/r_0\), and rewrite each density profile in Table \ref{Tab:Den-Valores} in the form \(\rho(r) = \rho_0 f(x)\). Then, performing the change of variables \(s = r_0 u\) gives
{\footnotesize
\[
\int_{r_*}^{r} s\,\rho(s)\,\mathrm ds
=
\rho_0 r_0^2
\int_{x_*}^{x} u f(u)\,\mathrm du
=
\rho_0 r_0^2
\left[\mathcal J(x)-\mathcal J(x_*)\right],
\]
}
where \(\mathcal J'(x) = x f(x)\).

For the Burkert, M-SFDM, and Einasto profiles, we have, respectively,
\[
\mathcal J_{\mathrm B}(x)
=
-\frac{1}{2}\ln(1+x)
+\frac{1}{4}\ln(1+x^2)
+\frac{1}{2}\arctan x,
\]
\[
\mathcal J_{\mathrm M}(x)
=
\frac{1}{2}\left[\ln x-\operatorname{Ci}(2x)\right],
\]
and
\[
\mathcal J_{\mathrm E}(x)
=
\frac{1}{\alpha}
\gamma\!\left(\frac{2}{\alpha},x^\alpha\right),
\]
where \(\gamma(a,z)\) denotes the lower incomplete gamma function, and \(\operatorname{Ci}(2x):=-\int_z^\infty \frac{\cos t}{t} \dd t\) .

As \(x\to\infty\), the Burkert and Einasto functions approach finite constants:
\[
\mathcal J_{\mathrm B}(\infty)=\frac{\pi}{4},
\qquad
\mathcal J_{\mathrm E}(\infty)
=
\frac{1}{\alpha}\Gamma\!\left(\frac{2}{\alpha}\right).
\]
In contrast, \(\mathcal J_{\mathrm M}(x)\) for M-SFDM exhibits logarithmic growth, so the associated estimate must be interpreted only within the finite halo range adopted in the numerical computations.

Introducing the dimensionless halo parameter
\[
\epsilon_h:=\frac{8\pi G\rho_0r_0^2}{c^2}.
\]
Since \(B(s)\leq B_{\max}\), the density contribution satisfies
{\footnotesize
\begin{equation}
\left|
\frac{8\pi G}{c^2}
\int_{r_*}^{r}B(s)s\varrho(s)\,\mathrm ds
\right|
\leq
B_{\max}\epsilon_h
\left|\mathcal J(x)-\mathcal J(x_*)\right|.
\label{eq:densidad prueba}
\end{equation}
}

Using the parameteres in Table \ref{Tab:Den-Valores}, we find
\[
\epsilon_h\simeq8.78\times10^{-7}
\quad\text{for Burkert},
\]
\[
\epsilon_h\simeq8.83\times10^{-7}
\quad\text{for M-SFDM},
\]
\[
\epsilon_h\simeq1.54\times10^{-6}
\quad\text{for Einasto}.
\]

Thus, the total integrated contribution to the density is of order \(10^{-6}\), up to the finite dimensionless profile factor in Eq. \eqref{eq:densidad prueba}, while the radial-pressure contribution is of higher order. Now let
\[
\mathcal M(x):=
\int_{x_{\mathrm{in}}}^{x}u^2f(u)\,\mathrm du,
\qquad
x_{\mathrm{in}}:=\frac{r_{\mathrm{in}}}{r_0},
\]
then \( m_h(r)=4\pi\rho_0r_0^3\mathcal M(x) \), and therefore \(q(r)=\frac{Gm_h(r)}{c^2r} = \frac{\epsilon_h}{2} \frac{\mathcal M(x)}{x}\). On any finite radial interval, \(\mathcal M(x)/x\) remains bounded, which implies \(q(r)=\mathcal O(\epsilon_h)\). In more detail,
{\footnotesize
\begin{equation}
4\left|
\int_{r_*}^{r}B(s)q^2(s)\frac{\mathrm ds}{s}
\right|
\leq
4B_{\max}q_{\max}^2
\ln\!\left(\frac{r}{r_*}\right)
=
\mathcal O(\epsilon_h^2).
\label{eq:presion prueba}
\end{equation}
}
The contribution from pressure is so small that it can be neglected.

Combining Eqs \eqref{eq:AB-integral prueba}, \eqref{eq:densidad prueba}, and \eqref{eq:presion prueba}, we conclude that \(A(r)B(r)=1+\mathcal O(10^{-6}) \), where \(A(r_*)B(r_*)=1\), and \(\varepsilon_h \approx 10^{-6}\). Thus
\begin{equation}
A(r)
=
\frac{1}{B(r)}
\left[1+\mathcal O(10^{-6})\right].
\label{eq:A-inverse-B-proof}
\end{equation}

The approximate reciprocal relationship between the metric functions is thus not an extra assumption. Instead, it arises from the small, dimensionless parameter \(\epsilon_h\) that characterizes the dark-matter halo. The dominant correction comes from the density, while the term linked to the reconstructed radial pressure appears only at order \(\epsilon_h^2\).

\begin{figure*}[h]
  \includegraphics[width=1.0\textwidth]{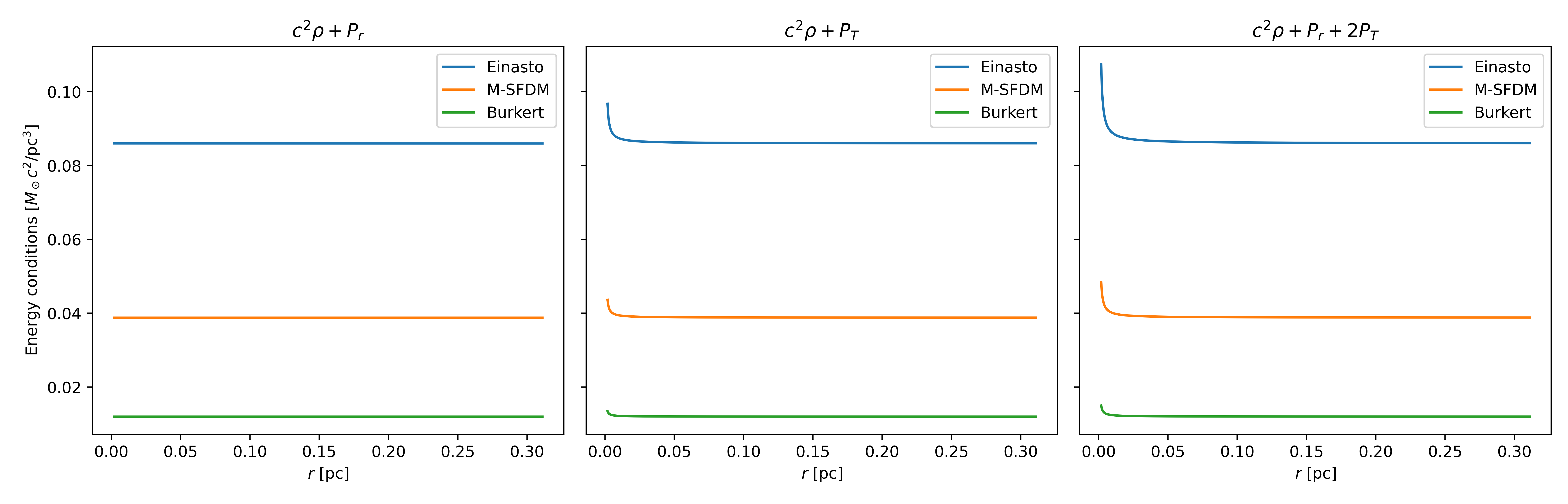}
\caption{Analysis of the strong energy condition for the Einasto, M-SFDM, and Burkert density profiles.}
\label{fig:energyconditions}
\end{figure*}

The use of a sharp Heaviside cutoff is not crucial for establishing the estimate \eqref{eq:A-inverse-B-proof}. In fact, it may be replaced by a one-sided smooth taper, \(\rho_\delta(r):=\rho_h(r)S_\delta(r)\), where \(0\leq S_\delta(r)\leq1\), \(S_\delta(r)=0\) for \(r\leq r_{\mathrm{in}}\), and \(S_\delta(r)=1\) for \(r\geq r_{\mathrm{in}}+\delta\). This modification smooths the halo’s onset while keeping the inner Schwarzschild vacuum region exactly intact. The associated mass function,
\[
m_{h,\delta}(r):=4\pi\int_{r_{\mathrm{in}}}^{r}s^2\rho_\delta(s)\,\mathrm ds,
\]
satisfies \(0\leq m_{h,\delta}(r)\leq m_h(r)\), which in turn implies \(q_\delta(r)\leq q(r)\) and \(B_\delta(r)\leq B(r)\). Hence, for all \(s\in [r_*,r]\), the two integrands in \eqref{eq:AB-integral prueba} satisfy
\[
B_\delta(s)s\rho_\delta(s)\leq B(s)s\rho(s),\qquad
B_\delta(s)q_\delta^2(s)\leq B(s)q^2(s).
\]
Accordingly, the introduction of a smooth transition can only decrease the upper bounds obtained in \eqref{eq:densidad prueba} and \eqref{eq:presion prueba}: the density term remains of order \(\epsilon_h\), and the radial-pressure term remains of order \(\epsilon_h^2\). Consequently, with the same temporal normalization, we still have
{\footnotesize
\[
A_\delta(r)B_\delta(r)=1+\mathcal O(\epsilon_h)=1+\mathcal O(10^{-6}),
\]
}
which demonstrates that, to leading order, the approximate reciprocal relation is insensitive to whether the inner halo edge is modeled as sharp or smoothly tapered.

\subsection{Strong energy condition}
\label{sec:3.3}

We will check whether the dark matter model satisfies the strong energy condition
\begin{equation*}
    T_{\mu\nu}t^\mu t^\nu+\frac{1}{2}T\geq0,
\end{equation*}
for any timelike unit vector $t^\mu$. The condition for the energy-momentum tensor of the anisotropic fluid \eqref{eq:T_Anisotropico} can be expressed as
\begin{equation*}
    \begin{split}
        c^2\rho+P_r(r)&\geq0, \\
        c^2\rho+P_T(r)&\geq0, \\
        c^2\rho+P_r(r)+2P_T(r)&\geq0.
    \end{split}
\end{equation*}
Figure \ref{fig:energyconditions} shows that each density profile satisfies the three strong energy conditions using the parameters shown in the Table \ref{Tab:Den-Valores}. The first two plots from left to right represent the radial and tangential energy conditions, respectively, while the final plot depicts the strong energy condition. All curves remain finite at the origin, since we assume that the density profile is nonzero only for $r>r_{in}$.

\subsection{Geometrical structure of the solution}
\label{SubSec:Geometrical_structure}

Introducing the inner radius \(r_{\mathrm{in}}\) splits spacetime into an inner vacuum region and an outer region filled with the dark matter halo. In the strong field region near the black hole, \(r<r_{\mathrm{in}}\), the halo mass and stress energy tensor vanish, \(m_h(r)=0,\ \rho(r)=P_r(r)=P_T(r)=0\), so Eq. \eqref{eq:psi-pr-general} reduces to \(\psi'(r)=0\), implying that \(\psi(r)\) is constant there. Denoting this constant by \(\psi_{\mathrm{in}}\), Eqs. \eqref{eq:B-with-halo} and \eqref{eq:A-ansatz} yield \(B(r)=\left( 1-\frac{2GM_\bullet}{c^2r} \right)^{-1}\) and \(A(r) = \left( 1-\frac{2GM_\bullet}{c^2r}\right)e^{2\psi_{\mathrm{in}}/c^2}\), and therefore,
\begin{equation}
A(r)B(r)
=
e^{2\psi_{\mathrm{in}}/c^2}
=
\mathrm{constant},
\qquad
r<r_{\mathrm{in}}.
\label{eq:AB_inner_region}
\end{equation}

Because the exponential factor is a constant, it can be locally absorbed by a constant rescaling of the time coordinate, \(\mathrm d\bar t = e^{\psi_{\mathrm{in}}/c^2}\mathrm dt\). Expressed in terms of \(\bar t\), the metric in the inner region assumes the Schwarzschild form. Consequently, the geometry for \(r<r_{\mathrm{in}}\) is locally isometric to Schwarzschild spacetime, consistent with Birkhoff’s theorem. However, once the time coordinate is fixed by matching it to the exterior region, the factor \(e^{2\psi_{\mathrm{in}}/c^2}\) can no longer be eliminated on a global scale. It then encodes a uniform gravitational redshift throughout the inner vacuum region, produced by the surrounding halo.

The Schwarzschild Killing horizon is located at \(r=\frac{2GM_\bullet}{c^2}\), by construction,
\[
r_h:=\frac{2GM_\bullet}{c^2}
<
r_{\mathrm{in}}
=
\frac{6GM_\bullet}{c^2},
\]
so the horizon is completely contained within the vacuum region. At this surface one has \(A(r_h)=0, \quad B^{-1}(r_h)=0\), while their product stays finite and positive:
\[
\lim_{r\to r_h}
A(r)B(r)
=
e^{2\psi_{\mathrm{in}}/c^2}.
\]
Therefore, the divergence of \(B(r)\) at this radius is the usual Schwarzschild coordinate singularity.

This conclusion is confirmed by the curvature invariants. Since the inner geometry is locally Schwarzschild, \(R=0, \quad R_{\mu\nu}R^{\mu\nu}=0\),  while the Kretschmann scalar is \(R_{\mu\nu\alpha\beta}R^{\mu\nu\alpha\beta}=\frac{48G^2M_\bullet^2}{c^4r^6}\). This invariant remains finite at \(r=2GM_\bullet/c^2\). Hence, the dark matter halo does not change the coordinate position or the intrinsic area of the Schwarzschild horizon. 

We now verify that no additional horizon is generated in the region occupied by the halo. For \(r>r_{\mathrm{in}}\), and using \eqref{eq:Velocidad Circular}, the radial metric function \eqref{eq:B-ansatz}, can be written as \( B^{-1}(r)=1-\frac{2GM_\bullet}{c^2r}-2q(r)\), since \(r>r_{\mathrm{in}}\), then \(\frac{2GM_\bullet}{c^2r} < \frac{1}{3}\), and, in the weakly relativistic halo regime, \(q(r)=\frac{v_c^2(r)}{c^2}\ll1\), it follows that \(B^{-1}(r)>\frac{2}{3}-2q(r)>0,\quad r>r_{\mathrm{in}}\), thus, \(B(r)\) remains finite throughout the finite halo domain under consideration.

The temporal metric function is strictly positive in this region: \(A(r)=\left(1-\frac{2GM_\bullet}{c^2r}\right)e^{2\psi(r)/c^2}>0\) for \(r>r_{\mathrm{in}}\). Therefore, \(A(r)>0,\quad B^{-1}(r)>0\) for \(r>r_{\mathrm{in}}\), and the dark matter halo does not generate an additional Killing horizon.
%%%%%%%

\paragraph{Regular matching at the inner halo boundary.}

It remains to verify that the transition between the inner Schwarzschild region and the exterior dark-matter halo at \(r=r_{\mathrm{in}}\) satisfies the Darmois-Israel junction conditions without generating a thin shell.
The density is \(\rho(r)=\Theta(r-r_{\mathrm{in}})\,\rho_h(r)\), so the halo mass is given explicitly by \eqref{eq:mh_definition}. For \(r<r_{\mathrm{in}}\), reversing the integral shows its integrand vanishes almost everywhere because \(\Theta(\tilde r-r_{\mathrm{in}})=0\) for \(\tilde r<r_{\mathrm{in}}\). At \(r=r_{\mathrm{in}}\), the integral is zero since the integration interval has zero length, so the value of \(\Theta(0)\) is irrelevant.

For the density profiles considered here, \(\rho_h(r)\) is finite near \(r=r_{\mathrm{in}}\). Hence there exist positive constants \(C\) and \(\delta\) such that
\( \left|\varrho_h(r)\right|\leq C,\quad r_{\mathrm{in}}\leq r\leq r_{\mathrm{in}}+\delta.\)
For \(r_{\mathrm{in}}<r<r_{\mathrm{in}}+\delta\), \eqref{eq:mh_definition} gives
\[
    \left|m_h(r)\right| \leq
    4\pi C
    \int_{r_{\mathrm{in}}}^{r}
    \tilde r^{2}
    \,\mathrm d\tilde r
    =
    \frac{4\pi C}{3}
    \left(
    r^3-r_{\mathrm{in}}^3
    \right),
\] since \(r^3-r_{\mathrm{in}}^3\to0\) as \(r\to r_{\mathrm{in}}^+\), the squeeze theorem implies 
\[
    \lim_{r\to r_{\mathrm{in}}^+}m_h(r)=0.
\]
Because \(m_h(r)=0\) throughout the inner region, we also have
\[
    \lim_{r\to r_{\mathrm{in}}^-}m_h(r)
    =
    \lim_{r\to r_{\mathrm{in}}^+}m_h(r)
    =
    0.
\]
Thus, the halo mass is continuous at the inner boundary. More precisely,
\[
m_h(r)
=
\mathcal O\!\left(r-r_{\mathrm{in}}\right)
\qquad
\text{as}
\qquad
r\to r_{\mathrm{in}}^+.
\]

Since the integrand in Eq.~\eqref{eq:mh_definition} is locally integrable, \(m_h(r)\) is locally absolutely continuous. Its distributional derivative is thus given by \eqref{eq:mprima} without any term proportional to \(\delta(r-r_{\mathrm{in}})\). Although \(m_h'(r)\) can have a finite jump at the inner boundary, \(m_h(r)\) itself is continuous, and no mass is concentrated on the matching surface.

We now examine the metric functions. For \(r<r_{\mathrm{in}}\), \eqref{eq:B-ansatz} reduces to
\(
B_-(r)
=
\left(
1-\frac{2GM_\bullet}{c^2r}
\right)^{-1},
\)
whereas for \(r>r_{\mathrm{in}}\),
\(
B_+(r)
=
\left(
1-\frac{2GM_\bullet}{c^2r}
-\frac{2Gm_h(r)}{c^2r}
\right)^{-1}.
\)
Using \eqref{eq:mh_definition}, it follows that
\[
\lim_{r\to r_{\mathrm{in}}^-}B_-(r)
=
\lim_{r\to r_{\mathrm{in}}^+}B_+(r)
=
\left(
1-\frac{2GM_\bullet}{c^2r_{\mathrm{in}}}
\right)^{-1}.
\]
Since \(r_{\mathrm{in}}=6GM_\bullet/c^2\), the common value is
\[
\lim_{r\to r_{\mathrm{in}}^\pm}B(r)
=
\frac{3}{2}.
\]

The explicit radial pressure is given by  \eqref{eq:Pr-explicito para prueba}, since \(m_h(r)=\mathcal O(r-r_{\mathrm{in}})\), we obtain
\(
P_r(r)
=
\mathcal O\!\left[
\left(r-r_{\mathrm{in}}\right)^2
\right],
\) and hence
\[
\lim_{r\to r_{\mathrm{in}}^+}P_r(r)=0.
\]

In the interior vacuum region, where \(m_h(r)=P_r(r)=0\), equation \eqref{eq:Pt_from_TOV} yields
\(
\psi'(r)=0,
\quad
r<r_{\mathrm{in}},
\)
on the halo side, the same equation becomes \eqref{eq:Pt_from_TOV}. Near \(r=r_{\mathrm{in}}\), the pressure term in the numerator is \(\mathcal O[(r-r_{\mathrm{in}})^2]\), while the term proportional to \(m_h(r)\) is \(\mathcal O(r-r_{\mathrm{in}})\). Thus, the full numerator is \(\mathcal O(r-r_{\mathrm{in}})\). On the other hand,
\begin{align*}
    \lim_{r\to r_{\mathrm{in}}^+}
    \left[
    c^2r-2GM_\bullet-2Gm_h(r)
    \right]
    &=
    c^2r_{\mathrm{in}}-2GM_\bullet
    \nonumber\\
    &=
    4GM_\bullet
    \neq0.
\end{align*}
Consequently,
\(
    \psi'(r)
    =
    \mathcal O\!\left(r-r_{\mathrm{in}}\right),
    \quad
    r\to r_{\mathrm{in}}^+,
\)
and
\[
    \lim_{r\to r_{\mathrm{in}}^-}\psi'(r)
    =
    \lim_{r\to r_{\mathrm{in}}^+}\psi'(r)
    =
    0.
\]

The inner solution determines \(\psi(r)\) only up to a constant. This constant is fixed by requiring continuity with the exterior solution 
\(
\lim_{r\to r_{\mathrm{in}}^-}\psi(r)
=
\lim_{r\to r_{\mathrm{in}}^+}\psi(r).
\)
It follows from \eqref{eq:A-ansatz} that
\[
\lim_{r\to r_{\mathrm{in}}^-}A(r)
=
\lim_{r\to r_{\mathrm{in}}^+}A(r).
\]

Furthermore, using Eq.~\eqref{eq:Aprime-over-A-final} together with the fact of \(\psi\)-continuity, we find
\[
    \lim_{r\to r_{\mathrm{in}}^\pm}
    \frac{A'(r)}{A(r)}
    =
    \frac{2GM_\bullet}
    {r_{\mathrm{in}}
    \left(
    c^2r_{\mathrm{in}}-2GM_\bullet
    \right)}=
    \frac{1}{2r_{\mathrm{in}}}.
\]
Because \(A(r)\) is continuous and nonzero at
\(r=r_{\mathrm{in}}\), this also implies
\[
\lim_{r\to r_{\mathrm{in}}^-}A'(r)
=
\lim_{r\to r_{\mathrm{in}}^+}A'(r).
\]

We now apply the Darmois-Israel junction conditions. The hypersurface \(\Sigma\colon r=r_{\mathrm{in}}\) is timelike because
\(
A(r_{\mathrm{in}})>0,
\qquad
B(r_{\mathrm{in}})=\frac{3}{2}>0.
\)
Its induced line element is
\[
\left.\mathrm ds^2\right|_{\Sigma}
=
-A(r_{\mathrm{in}})c^2\,\mathrm dt^2
+
r_{\mathrm{in}}^2
\left(
\mathrm d\theta^2
+
\sin^2\theta\,\mathrm d\phi^2
\right).
\]
The continuity of \(A \) shows that the induced metric is continuous across the matching surface.

Using a normal vector directed toward increasing \(r\) on both sides,
\[
n^\mu
=
\left(
0,
\frac{1}{\sqrt{B(r)}},
0,
0
\right),
\]
and the convention
\(
    K_{ab}
    =
    \frac{1}{2}\mathcal L_n h_{ab},
\)
, \(\mathcal L_n\) denotes the Lie derivative along the radial vector \(n\), and \(h_{ab}\) is the metric; the nonzero mixed components of the extrinsic curvature are
\begin{equation}
K^t{}_t
=
\frac{A'(r)}
{2A(r)\sqrt{B(r)}},
\qquad
K^\theta{}_\theta
=
K^\phi{}_\phi
=
\frac{1}{r\sqrt{B(r)}}.
\label{eq:extrinsic_curvature_matching}
\end{equation}

Using the continuity of \(A,B\), the temporal component satisfies
\[
\lim_{r\to r_{\mathrm{in}}^\pm}
K^t{}_t
=
\frac{1}{4r_{\mathrm{in}}}
\sqrt{\frac{2}{3}},
\]
while the angular components satisfy
\[
\lim_{r\to r_{\mathrm{in}}^\pm}
K^\theta{}_\theta
=
\lim_{r\to r_{\mathrm{in}}^\pm}
K^\phi{}_\phi
=
\frac{1}{r_{\mathrm{in}}}
\sqrt{\frac{2}{3}}.
\]
Therefore, every component of the extrinsic curvature has the same one-sided limit:
\begin{equation}
\left[K^a{}_b\right]=0.
\label{eq:K_jump_zero_matching}
\end{equation}
Here,
\(
    \left[K^a{}_b\right]
    :=
    \lim_{r\to r_{\mathrm{in}}^+}K^a{}_b
    -
    \lim_{r\to r_{\mathrm{in}}^-}K^a{}_b.
\)

The Israel surface stress-energy tensor is
\[
S^a{}_b
=
-\frac{c^4}{8\pi G}
\left(
\left[K^a{}_b\right]
-
\delta^a{}_b[K]
\right).
\]
Since
\(
\left[K^a{}_b\right]=0,
\qquad
[K]=0,
\)
we obtain
\begin{equation}
S^a{}_b=0.
\label{eq:Israel_tensor_zero}
\end{equation}

Thus, the inner Schwarzschild region and the outer dark-matter halo are joined by a regular matching at \(r=r_{\mathrm{in}}\). No thin shell, surface energy density, surface pressure, or surface tension is present. The density \(\varrho(r)\) may have a finite jump at the halo onset, causing finite discontinuities in \(m_h'(r)\), higher metric derivatives, and curvature, but the induced metric and extrinsic curvature remain continuous. Hence no Dirac-delta terms arise in the curvature or stress-energy tensor. The inner boundary marks the regular start of the extended dark-matter distribution rather than an extra material shell, and no surface energy conditions are violated.

\section{Results}
\label{sec:4}

To describe the dark matter halo, we will use the density profiles shown in Table \ref{Tab:Den-Valores}, along with the characteristic values of the parameters $\rho_0$ and $r_0$ also shown there. We will focus on observational data from star rotation curves in the Milky Way.

\begin{table}[ht]
\centering
\caption{Density profiles obtained from \cite{Cabrera2023}. Values $\rho_0$ and $r_0$ for Milky Way galaxy obtained from Burkert \cite{lin2019}, M-SFDM \cite{Bernal:2024}, Einasto \cite{ou2024}. For the Einasto profile, we also have the parameter $\alpha=0.91^{+0.04}_{-0.05}$.}
\label{Tab:Den-Valores}
\begin{tabular}{llll}
\hline\noalign{\smallskip}
\textbf{Profile} & $\rho_h(r)$ & $\rho_0 (10^{-2}\frac{M_\odot}{pc^3})$ & $r_0(kpc)$ \\ 
\noalign{\smallskip}\hline\noalign{\smallskip}
Burkert & $\rho_0\frac{r_0^3}{(r_0+r)(r_0^2+r^2)}$ & $1.2\pm1$ &  $7.8\pm0.4$ \\ 
M-SFDM & $\rho_0\frac{r_0^2}{r^2}\sin^2\left(\frac{r}{r_0}\right)$ & $3.88^{+1.16}_{-1.08}$ & $4.35^{+0.
57}_{-0.57}$ \\ 
Einasto & $\rho_0exp\left[-\left(\frac{r}{r_0}\right)^\alpha\right]$ & $8.6^{+3.0}_{-2.8}$ & $3.86^{+0.35}_{-0.38}$ \\
\noalign{\smallskip}\hline
\end{tabular} 
\end{table}

\begin{figure*}[h]
  \includegraphics[width=1.0\textwidth]{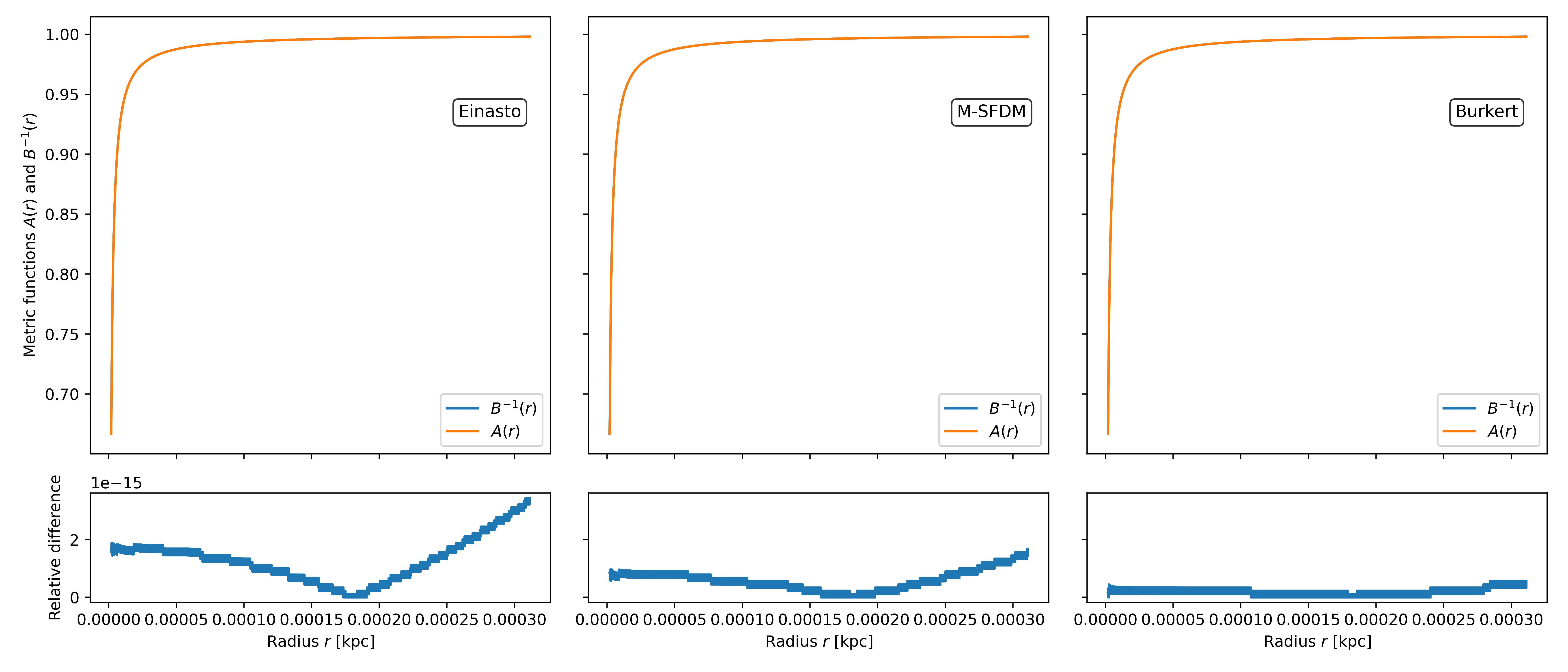}
\caption{Comparison of the metric functions $A(r)$ and $1/B(r)$ for each density profile and their relative difference.}
\label{fig:metricfunctioncompar}
\end{figure*}

\begin{figure}[h]
  \includegraphics[scale=0.35]{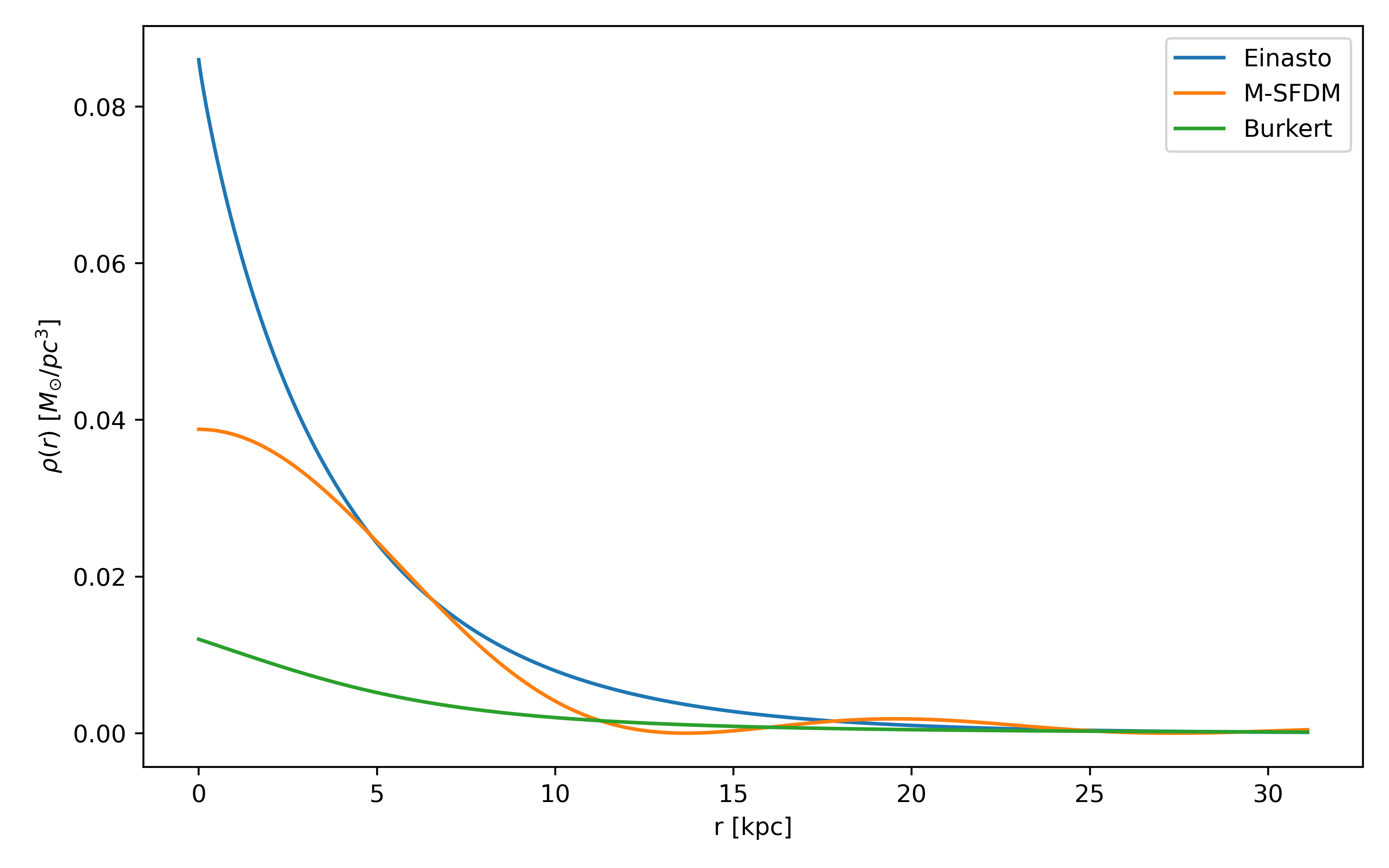}
\caption{Density profiles of the Einasto, M-SFDM, and Burkert models for the parameter values given in Tab.~\ref{Tab:Den-Valores}.}
\label{fig:densities}      
\end{figure}

Figure \ref{fig:densities} shows the density profiles used to model the dark matter halo. Since we are interested in the center of the galaxy, the M-SFDM model is treated only as an approximation to the ground state.

For the numerical analysis of the metric functions and the pressures and densities of dark matter, we will assume the mass of Sagittarius A* based on observational data from more than two dozen central stars, as given in \cite{Genzel2010}, which has a value of $M_\bullet=4\times 10^{6} M_\odot$.

%% \textcolor{blue}{\sout{Furthermore, we will assume that the density of dark matter is nonzero only if $r>r_{in}$, which is the boundary beyond which a geodesic of a massive particle is stable in the Schwarzschild black hole. This means that the contribution of dark matter to the system for $r<r_{in}$ is zero, because all the dark matter inside this radial threshold is captured by the black hole. This assumption is reflected in the definition of the dark matter density used in \eqref{eq:mprima}.}}

The metric functions $A(r)$ and $B^{-1}(r)$ are shown in Figure \ref{fig:metricfunctioncompar} for each of the density profiles. It can be seen that there are small differences among the three density profiles used even when they have different origins: phenomenological, simulation-based, or derived from a theoretical model, respectively.

Furthermore, we can see that both metric functions converge to Minkowski spacetime, that is $A(r)=B^{-1}(r)=1$, as we move away from the center of the galaxy, this result supports the Newtonian approximations used in previous studies, and is to be expected given the distance from the black hole, which allows its gravitational effects to be neglected.

The most important result of this numerical and semi-analytical analysis is the fact that the metric functions, even when no relationship between them has been assumed, exhibit the behavior of the vacuum solution $A(r) \approx B^{-1}(r)$. This equality has been adopted in previous works from the outset without a solid justification such as that presented in this paper. This paves the way for future work in which this relationship is imposed from the outset, not merely as an ansatz but as a previously proven result.

\begin{figure}[h]
  \includegraphics[scale=0.4]{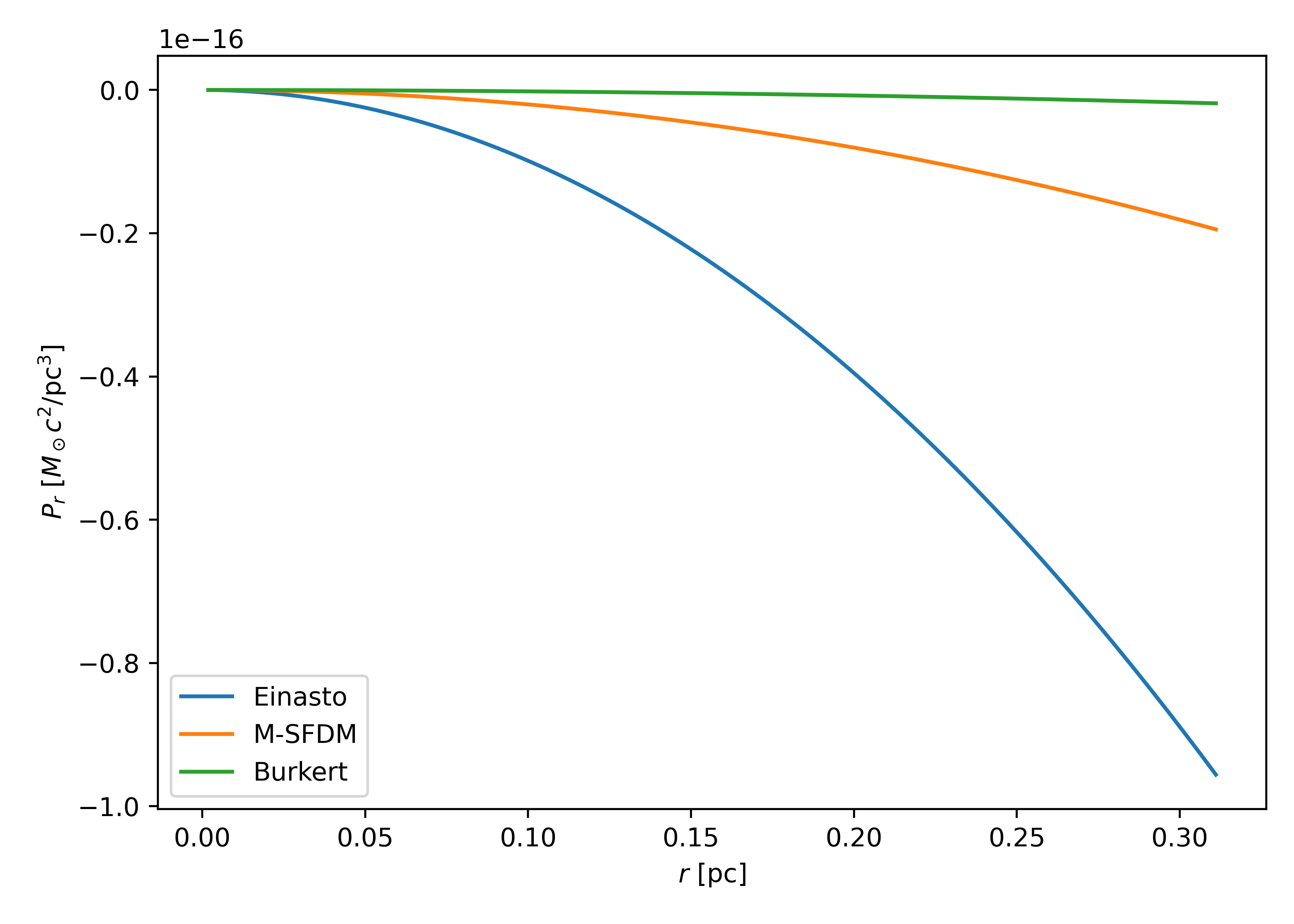}
\caption{Comparison of radial pressure for each density profile.}
\label{fig:Prcomparation}      
\end{figure}

\begin{figure}[h]
  \includegraphics[scale=0.4]{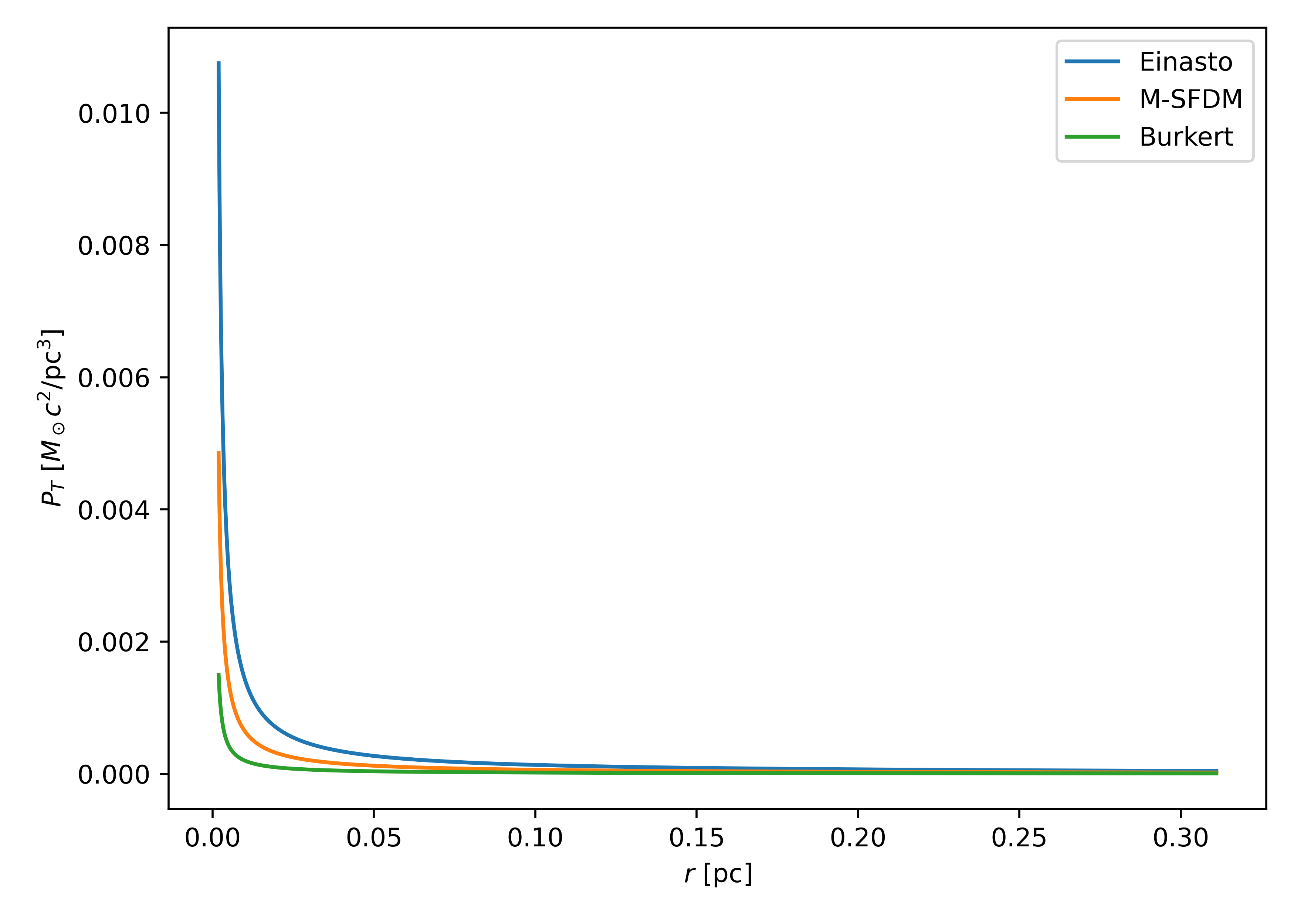}
\caption{Comparison of tangential pressure for each density profile.}
\label{fig:Ptcomparation}      
\end{figure}

\begin{figure*}
  \includegraphics[width=1.0\textwidth]{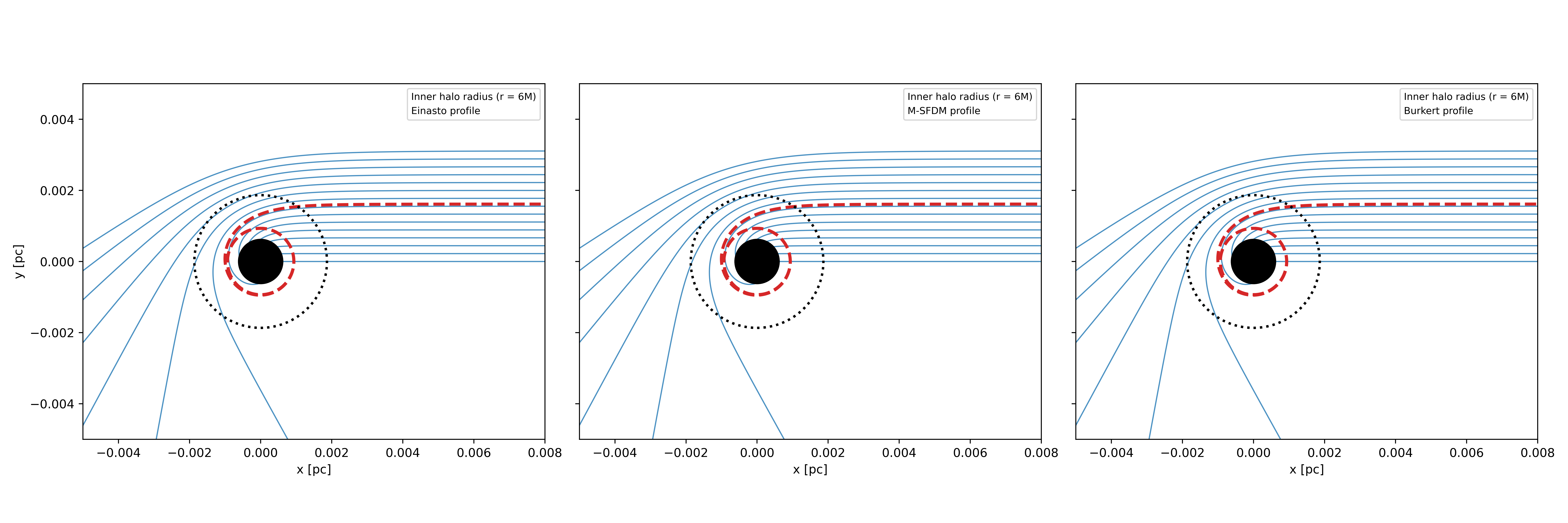}
\caption{Null geodesics for each density profile.}
\label{fig:geodesics}
\end{figure*}

This behavior does not arise from a dynamical compensation between the radial and tangential pressures, which are shown in Figures \ref{fig:Prcomparation} and \ref{fig:Ptcomparation}. Instead, the Einstein equations imply that the variation of the product \(A(r)B(r)\) is governed by the combination \(c^2\varrho(r)+P_r(r)\). In the weakly relativistic regime considered here, the corresponding contribution is small, while the radial-pressure term provides only a higher-order correction in the compactness of the matter distribution. The tangential pressure does not enter directly into this relation. Consequently, the approximate reciprocal behavior
\(
A(r)\simeq B^{-1}(r)
\)
emerges from the small relativistic strength of the matter source, rather than from an adjustment of the pressure components or from an exact recovery of a vacuum geometry. The resulting spacetime therefore retains a vacuum-like reciprocal structure only up to small matter-induced corrections.

Lastly, we show the null geodesics in an equatorial plane in the figure \ref{fig:geodesics}. For each profile, the stable null geodesic around the black hole is marked with the red dashed line. The black dashed line represents the limit $r_{in}=\frac{6GM_\bullet}{c^2}$. Only when the distance from the black hole increases can we note the difference among the density profile used. Consequently, from the perspective of null geodesics, we cannot determine whether these black holes are embedded in dark matter halos or are isolated.

\subsection{Observational constraints}

\subsubsection{Pericenter precession for stars}

We now study the motion of a particle in spacetime as determined by the metric presented in this paper. Owing to the symmetries of the metric, the energy and angular momentum per unit mass are conserved and can be written as
\begin{equation*}
    E=A(r)c^2\dot t, \qquad L=r^2\dot\phi,
\end{equation*}
using the normalization condition for a timelike geodesic, $g_{\mu\nu}\dot{x}^\mu\dot{x}^\nu=-c^2$, and express it in terms of these conserved quantities, we obtain the radial equation
\begin{equation*}
    \dot{r}^2=\frac{1}{B(r)}\left(\frac{E^2}{A(r)c^2}-c^2-\frac{L^2}{r^2}\right),
\end{equation*}
given that $\dot{r}=\frac{L}{r^2}\frac{dr}{d\phi}$ and introducing the variable $u=1/r$, the radial equation can be recast as
\begin{equation*}
    \left(\frac{du}{d\phi}\right)^2 =\frac{1}{L^2B(u)}\left(\frac{E^2}{A(u)c^2}-c^2-u^2L^2\right),
\end{equation*}
to obtain a second-order differential equation, we can differentiate the last one and get
\begin{equation}\label{eq:edo_pericenter}
\begin{aligned}
\frac{d^2u}{d\phi^2}
={}&\frac{\dot{B}(u)}{2L^2B(u)^2}
\left(-\frac{E^2}{A(u)c^2}+c^2+u^2L^2\right)\\
&-\frac{1}{2L^2B(u)}
\left(\frac{E^2\dot{A}(u)}{A(u)^2c^2}+2uL^2\right).
\end{aligned}
\end{equation}
Equation \eqref{eq:edo_pericenter} is solved numerically to determine the orbital evolution and, in particular, the pericenter precession. To compare our results with the observed orbit of the S2 star, we first determine the pericenter and apocenter distances of the corresponding Newtonian ellipse,
\begin{equation*}
    r_p=a(1-e), \qquad r_a=a(1+e),
\end{equation*}
using the reported orbital parameters $a/r_o=125.058\pm0.041 mas$ and $e=0.884649\pm0.000066$ \cite{GRAVITY:2020gka,GRAVITY:2024tth}. The star S2 orbits near the center of the Milky Way at a radius $r_0=1400R_S$, and the precession of the pericenter per orbit is given by
\begin{equation*}
    \Delta\phi=f_{sp}\times12.1am,
\end{equation*}
where $f_{sp}$ is used as a fitting parameter, the optimal value obtained is $f_{sp}=1.11\pm 0.21$.

We integrated the differential equation \eqref{eq:edo_pericenter} over one complete orbit for each of the three density profiles considered in this work. In all three cases, we obtain $\Delta\phi=12.1687 am$, this value lies within the observational range inferred from the S2 orbit, indicating that the predicted pericenter precession is consistent with the measurements reported by the GRAVITY Collaboration.

\subsubsection{Shadow size of Sagittarius A*}

We now turn to the shadow size of the black hole using the results presented in \cite{Perlick:2021} for a general spherically symmetric and static metric. The angular radius of the shadow $\alpha_s$ is
\begin{equation}\label{eq:angular_size_shadow}
    \sin^2{\alpha_s}=\frac{h(r_{ph})^2}{h(r_o)^2},
\end{equation}
where $r_o$ is the observer radial coordinate, $r_{ph}$ is the radius of the photon sphere, and the function $h(r)$ is derived from the metric according to
\begin{equation*}
    h(r)^2=\frac{r^2}{A(r)}.
\end{equation*}
As shown previously, the metric considered in this work is asymptotically flat. Therefore, for an observer located at $r_o$, the function $h(r_o)$ reduces to $r_o$, and \eqref{eq:angular_size_shadow} can be approximated by
\begin{equation}\label{eq:angular_size_shadow_final}
    \alpha_s=\frac{h(r_{ph})}{r_o}.
\end{equation}
Once the photon-sphere radius $r_{ph}$ is determined, the angular radius of the shadow can be calculated from equation \eqref{eq:angular_size_shadow_final}. We take the distance to SgrA* to be $r_o=8178pc$ \cite{gravity2019} and compare our prediction with the observational measurement $2\alpha_s=51.8\pm2.3\mu as$ reported by the Event Horizon Telescope Collaboration \cite{EventHorizonTelescope:2022}. For the three density profiles considered in this work, we obtain $2\alpha_s=50.1746\mu as$, this result is consistent with the observational range reported by the Event Horizon Telescope Collaboration.

Therefore, the dark matter halo configurations considered in this work are compatible with the current observational constraints on both the S2 pericenter precession and the shadow size of SgrA*.

\section{Conclusions}
\label{sec:5}
We have presented an alternative method for analyzing the system consisting of a black hole surrounded by a dark matter halo, in which we ensure that the metric functions are solutions to the Einstein field equations without assuming any relationship between them, unlike the models used in recent studies.

The dark-matter halo has been modeled as an anisotropic fluid. This description allows the density profile to be prescribed independently, while the radial pressure is reconstructed from the tangential velocity relation and the tangential pressure is subsequently determined from the conservation equation. In contrast to previous work, this hypothesis regarding the halo model allows for greater flexibility regarding the nature of dark matter itself and does not focus on models in which $P_r=0$ or $P_r=P_T$, such as the CDM model.

The numerical and semi-analytical analyses lead to two main conclusions. First, although no relation between \(A(r)\) and \(B(r)\) is imposed in the metric ansatz, the resulting functions satisfy \(A(r)B(r)=1+\mathcal O(10^{-6})\) throughout the finite halo domain considered. This result does not arise because the radial and tangential pressures dynamically compensate for a mismatch between the metric functions.  Instead, the Einstein equations imply that the variation of the product \(A(r)B(r)\) is governed by the dark matter density and is controlled by the factor \(\varepsilon_h=\frac{8\pi G\rho_0r_0^2}{c^2}\). 

For the galactic parameters considered here, \(\varepsilon_h\sim10^{-6}\). The radial-pressure contribution is subleading because it is quadratic in the halo compactness, while the tangential pressure does not enter directly into the relation between \(A(r)\) and \(B(r)\).

Second, the three density profiles produce qualitatively similar metric functions. Their quantitative differences remain small because all three configurations are characterized by a weak relativistic halo strength. The function \(B(r)\) is determined directly by the enclosed halo mass, whereas \(A(r)\) is obtained from the halo induced deformation of the temporal sector. The close agreement between the resulting metric functions is therefore a consequence of the small dimensionless gravitational strength of the corresponding halo parameters, rather than a cancellation produced by the pressure components.

In addition, the inner Schwarzschild region and the outer dark matter halo are regularly matched at $r = r_{\text{in}}$. The fulfillment of the Darmois–Israel junction conditions guarantees the regularity of the transition and completes the global spacetime description obtained in this study.

Finally, we have tested the observational implications of the resulting spacetime using two independent observables associated with SgrA*. We calculated the pericenter precession of the S2 star and obtained $\Delta\phi=12.1687 am$ for all three density profiles considered. This value lies within the observational range reported by the GRAVITY Collaboration. We also calculated the angular diameter of the shadow of Sgr~A*, obtaining $2\alpha_s=50.1746\mu as$ again for all three density profiles. This prediction is consistent with the Event Horizon Telescope measurement.

Therefore, the model satisfies two independent observational constraints: the orbital dynamics of the S2 star and the shadow size of SgrA*. The agreement with both measurements indicates that the presence of the dark matter halo, within the parameter ranges considered here, does not produce deviations incompatible with current observations.

In summary, our results provide a theoretical basis for the relation $A(r)B(r)\simeq1$ in black hole-dark matter halo systems, rather than treating it solely as an initial assumption. The analysis is applicable to density profiles of different origins, including phenomenological models, theoretical constructions, and profiles derived from numerical simulations. This provides a useful framework for future investigations in which the relationship between the metric functions and the dark matter equation of state can be studied using the relation obtained here as a derived result rather than as an imposed ansatz.

\begin{acknowledgements}
This work was partially supported by SECIHTI Mexico under grants CBF-2025-G-1720 and CBF-2025-G-176. Also by the grant I0101/131/07 C-234/07 of the Instituto Avanzado de Cosmolog\'ia (IAC) Collaboration (http://www.iac.edu.mx/), and for the computing time granted by LANCAD and SECIHTI in the Supercomputer Hybrid Cluster ``Xiuhcoatl" at the General Coordination of Technological Services in Information and Communications (CGSTIC) of CINVESTAV, IPN (http://clusterhibrido.cinvestav.mx).
\end{acknowledgements}
\bibliographystyle{spphys}
\bibliography{References}   

\end{document}